\documentclass[onecolumn,12pt]{article}  
\usepackage{amssymb}
\usepackage{graphicx}
\usepackage{dcolumn}
\usepackage{bm}
\usepackage{epsfig}
\usepackage{hyperref}

\RequirePackage{graphicx}

\newcommand{\co}[1]{#1}
\newcommand{\cu}[1]{#1}

\begin{document}

\title{Thermodynamics of interacting holographic dark energy}

\author{Fabiola Arevalo$^1$\footnote{fabiola.arevalo@ufrontera.cl}, Paulo Cifuentes$^1$\footnote{p.cifuentes@ufromail.cl} and Francisco Pe\~{n}a$^1$\footnote{francisco.pena@ufrontera.cl}\\ \small{
$^1$ Departamento de Ciencias F\'isicas, Facultad de Ingenier\'ia y Ciencias, }\\
\small{Universidad de La Frontera, Temuco, Casilla 54-D, Chile }\\
}

\date{\today }
\maketitle
\begin{abstract}
The thermodynamics of a scheme of dark matter-dark energy interaction is studied considering a holographic model for the dark energy in a flat Friedmann-Lemaitre-Robertson-Walker background. We obtain a total entropy rate for a general horizon and we study the Generalized Second Law of Thermodynamics for a cosmological interaction as a free function. Additionally, we discuss two horizons related to the Ricci and Ricci-like model and its effect on an interacting system.
\end{abstract}

\maketitle

\section{Introduction}\label{intro}
Observations of type Ia Supernova \cite{Perlmutter:1998np,Riess:1998cb} and the cosmic microwave background \cite{Komatsu:2010fb}, among other more recent data \cite{Ade:2013zuv}, strongly suggest that the universe is in a phase of late accelerated expansion. The most accepted interpretation in the context of Einstein's General Relativity is that this phase is driven by an unknown component called dark energy \cite{Copeland:2006wr}, which would account for around $73\%$ of the total energy density in the universe today.
This component is usually described as a fluid with negative pressure. Various theoretical models have been proposed to account for it. One such model, the holographic dark energy model, is based on the application of the holographic principle to cosmology. 
According to \cite{Cohen:1998zx} the energy contained in a region of size $L$ must not exceed the mass of a black hole of the same size, which means, in terms of energy density, $\rho\leq
L^{-2}$. If in a cosmological context it is considered $L=H^{-1}$, where $H$ is the Hubble parameter, then $\rho_{_{DE}}\propto H^{2}$, where $\rho_{_{DE}}$ is the dark energy density, giving a model of holographic nature for this density. Based on this idea, M. Li \cite{Li:2004rb} proposed the model $\rho_{_{DE}}=3{c}^{2}{H^{2}}$, known as Holographic Dark Energy (HDE).
Later on, the authors in \cite{Gao:2007ep} proposed $\rho_{_{DE}} \propto R$, where
$R=6\left(  {2{H^{2}}+\dot{H}+k/{a^{2}}}\right)$ is the Ricci scalar, model referred to as Ricci Holographic Dark Energy (RHDE). %X
In  \cite{Granda:2008dk}, inspired by RHDE for a flat model with $\rho_{_{DE}}=3\left(  {\alpha{H^{2}}+\beta \dot{H}%
}\right)$ was then proposed, called Ricci-like, where $\alpha$ and $\beta$ are constants. This has been extended to include the apparent and event horizon, an even a linear combination of both \cite{Sadjadi:2006qb}.
Models with interaction between dark energy and dark matter have been studied to explain the evolution of the Universe and the cosmic coincidence problem; modeled by a phenomenological function $Q$ describing energy transfer between two components.
Interacting models for HDE have been studied by \cite{Hu:2006ar},\cite{Chimento:2011pk} for Modified Holographic Ricci Dark Energy (MHRDE) and later on focusing on sign change interaction for a Ricci-like model \cite{Arevalo:2013tta}.\\
A change of sign in $Q$ has been proposed by \cite{Cai:2009ht}, where without choosing an interaction they found that the functional changes sign fitting with data. Following this line of research, the authors \cite{Li:2011ga} proposed a variable phenomenological interaction, where the sign change appears at early times when fitted with data.\\ 
There are studies in which the interaction is proportional to the deceleration parameter, to obtain a change of sign in the interaction when the universe transits from a non-accelerated phase to an accelerated one 
\cite{Wei:2010cs,Bolotin:2013jpa,Khurshudyan:2013ohq,Forte:2013fua}.
There are no thermodynamics studies to the author's knowledge however, there are studies of HDE model 
with
 thermodynamics for a positive interaction using the Generalized Second Law (GSL). 
\\
In \cite{Wang:2007ak} the authors propose a thermodynamical description of interacting HDE and dark matter. 
On this basis, the authors of \cite{Setare:2008bb} generalize it for a non-flat Universe, and obtain a relation for $Q$ in terms of $H$. How the interaction behaves considering thermodynamical factors is of interest given its unknown nature. \\
In \cite{Sheykhi:2009zv} the author reviews GSL of interacting HDE with the apparent horizon and express the total entropy change as a function of a ratio between energy densities for three interactions. The author also asks whether thermodynamics in an accelerating universe can reveal some properties of dark energy itself. 
In \cite{Karami:2009yd} the authors study GSL for interacting dark energy in a non-flat FRW universe enclosed by the dynamical apparent horizon, and expressing the total change in entropy as a function of variables estimated today  they conclude that the GSL is respected for the present time.  
Later on, the author in \cite{Debnath:2010za} generalizes this to three fluids 
 for an event horizon and concludes that the GSL is not respected for the present time.  
In \cite{Bhattacharya:2010zf} the authors test thermodynamical laws considering a total energy density to obtain the change in entropy and then using three variable EoS of HDE. A graphical analysis is performed (analytic expressions are not conclusive) and for a specific range one of the models respects GSL.\\
In \cite{Karami:2010zz} the authors study GSL at the apparent and event horizons; in the latter they establish an upper limit for the dark energy EoS.
 In \cite{Bhattacharya:2011} the authors study Generalized Holographic and RDE at the apparent horizon, particle horizon and event horizon. They use graphical investigation and find that the GSL cannot be satisfied at these horizon for this model.
In \cite{Mazumder:2011} they study HDE focusing on the State Finder Parameters with several horizons for a given interaction and finally in \cite{Mathew:2013fka} the authors consider interacting RHDE and GSL for the event horizon. \\
We follow the work of \cite{Setare:2007at,Jamil:2009eb} for three interacting fluids and the GSL for the apparent horizon. 
The present study was developed in the framework of a flat Friedmann-Lemaitre-Robertson-Walker (FLRW) cosmology with two interacting fluids, dark energy and dark matter. The main goal of this work is to examine the validity of GSL for an interacting holographic context bounded by a general surface, considering the holography Ansatz related to the horizon itself and study its effect on the sign-change cosmological interaction. We will also investigate obtaining limits on the parameters of our model from thermodynamics criteria.\\
This article is organized as follows: Section $2$ describes interacting fluids and their thermodynamics. Section $3$ studies the dark energy related to a length scale $L$ and later develop it using specific horizon for HDE. Finally, Section $4$ is devoted to discussion of our results.

\section{Interaction and the Generalized Second Law} \label{dos}
We begin our analysis considering a flat model with two fluids, $\rho_{_{DM}}$, the dark matter component, $p_{_{DM}}$ the dark matter pressure, $\rho_{_{DE}}$, the dark energy component with $p_{_{DE}}$ its pressure. We neglect other components given that we are interested in late Universe dynamics. The field equations, considering natural units throughout
this work, are 
\begin{eqnarray}
 3H^2=\rho_{_{DE}}+\rho_{_{DM}} ,\label{00}\\
 \dot H  =  - \frac{ 1}{2}\left({\rho _{_{DM}}} + {\rho _{_{DE}}} + p_{_{DE}}+p_{_{DM}}\right) , \label{yy}
\end{eqnarray}
where $H(a)=\dot a/ a$ is the Hubble parameter, $a$ is the scale factor and $\dot a$ denotes derivative of $a$ with respect to cosmic time $t$. The total energy density is $\rho(a)={\rho _{_{DM}}} + {\rho _{_{DE}}}$ and its equation of conservation is given as
\begin{eqnarray}
\dot{\rho}_{_{DE}}+\dot{\rho}_{_{DM}}+3H \left(\rho_{_{DE}}+\rho_{_{DM}}+p_{_{DM}}+p_{_{DE}}\right)=0
. \label{DE+DM}
\end{eqnarray}
Assuming that in our model, dark energy interacts with dark matter through a phenomenological coupling function denoted by $Q(a)$, we split the conservation equation into
\begin{eqnarray}
\dot{\rho}_{_{DE}}+3H\left(\rho_{_{DE}}+p_{_{DE}}\right)=-Q , \label{DE}\\
\dot{\rho}_{_{DM}}+3H \left(\rho_{_{DM}}+p_{_{DM}} \right)=Q .\label{DM}
\end{eqnarray}
The nature of the cosmological interaction between dark matter and dark energy is of phenomenological origin and its nature remains unknown \cite{Bertolami:2007zm,Costa:2009mv,Costa:2013sva,Xu:2013jma}.
Several model have been considered in the literature where $Q$ is a function of the energy densities and the Hubble parameter \cite{Chen:2011cy}, considering that equations (\ref{DE}) and (\ref{DM}) can be integrated straightforwardly \cite{Chimento:2009hj}. In the present study, we do not assume this coupling term \textit{a priori}, but rather study the system through the field equations and remain as general as possible as done by \cite{Cataldo:2010kc,Arevalo:2013tta}. 
We consider the coincidence parameter $r(a)$ as an auxiliary variable, defined as the ratio between the two energy densities
\begin{eqnarray}
r \equiv \frac{\rho_{_{DM}}}{\rho_{_{DE}}} \ .\label{r}
\end{eqnarray} 
This parameter is often used to address the cosmic coincidence problem: Why are energy densities of the same order of magnitude at late times? 
 The interaction function in terms of this parameter \cite{delCampo:2013hka} can be obtained replacing $\rho_{_{DE}}=\frac{3H^2}{1+r}$ in (\ref{DE}) \co{using (\ref{yy})}, then  
\begin{eqnarray}
\frac{Q}{3H^3}=\frac{1}{(1+r)^2} \left[r'-\frac{(r+1)}{H^2}\left( rp_{_{DE}} -p_{_{DM}}\right)\right], \label{Qpr}
\end{eqnarray}
where the interaction $Q$ is a free function which depends on $(H,r,p_{_{DE}},p_{_{DM}},r')$; the prime in $r'$ denotes a derivative of $\log a$. Given that the term with the pressures is negative and that the term $r'$ is negative at late times for a positive decreasing $r$ function, which would alleviate cosmic coincidence problem, this interaction could change sign, as noted by \cite{Arevalo:2013tta} for \co{pressureless dark matter and} a barotropic EoS in a holographic context. This model of late Universe will be studied considering thermodynamics and the possibility that the cosmological interaction could be a slightly negative function at some point during its evolution.

To examine the thermodynamic behavior of a cosmological scenario, the universe (in our case the interacting dark sector) must be considered as a thermodynamical system with a certain boundary. The Friedmann equations themselves arise from the first law of
thermodynamics, as shown by \cite{Padmanabhan:2009vy}. This follows the work of \cite{Jamil:2009eb}, where the laws of thermodynamics are valid and the late universe is interpreted as a state in thermodynamical equilibrium, this may not be the general case and studies where the components have different temperatures have been considered, see for instance \cite{Pavon:2007gt} and references therein. Entropy corrections have been considered in this scenarios linked to $Q$, however according to \cite{Wei:2009kp} this correction is relevant only at early times for the HDE. 
We will examine whether the sum of the entropy enclosed by the horizon 
and the entropy of the horizon 
 itself, denoted as $S_{tot}$, is not a decreasing function of time. This principle is known as the Generalized Second Law (GSL) of Thermodynamics and is expressed through the inequality $\dot S_{tot} \geq 0$. 

The first law of thermodynamics for the matter content is written as
\begin{eqnarray}
T_{_{DE}} d {S}_{_{DE}}&=& p_{_{DE}} dV+dE_{_{DE}}, \nonumber \\
T_{_{DM}} d{S}_{_{DM}} &=&p_{_{DM}}  dV +d E_{_{DM}} ,\label{s1}
\end{eqnarray}
where (${S}_{_{DE}}, {S}_{_{DM}}$) are the entropies of the dark components and ($E_{_{DE}},E_{_{DM}}$) their associated energies. The volume of the system $V = 4 \pi L^3 /3$ is bounded
by the radius $L$ and thus its differential form is $dV = 4 \pi L^2 d L$. 
The energy associated with each fluid is defined in relation to its energy density and the length scale as
\begin{eqnarray}
E_{_{DE}}\equiv \frac{4 \pi}{3} L^3 \rho_{_{DE}}, \quad E_{_{DM}}\equiv \frac{4 \pi}{3} L^3 \rho_{_{DM}}. \label{nueve}
\end{eqnarray}
Dividing the sum of (\ref{s1}) by $dt$ and considering the derivative of both equations in (\ref{nueve}), we obtain the entropy variation of the fluid inside the surface of radius $L$
\begin{eqnarray}
 \dot {S}_{_{DE}}+\dot {S}_{_{DM}}= \frac{4 \pi L^2}{T_{_{DE}}} \left( \dot L \left( p_{_{DE}}+\rho_{_{DE}} \right)+\frac{L}{3} \dot \rho_{_{DE}}\right) %&& \nonumber \\
+\frac{4 \pi L^2}{T_{_{DM}}} \left( \dot L \left( p_{_{DM}}+\rho_{_{DM}} \right)+\frac{L}{3} \dot \rho_{_{DM}}\right) &&
\label{s3}
\end{eqnarray}
In thermal equilibration, all the constituent fluids of the
universe are considered to have the same temperature $T$, while their energy
and pressure can, in general, be different. If we consider equilibrium for total energy density, as proposed by \cite{Jamil:2009eb} for holographic dark energy models, the temperatures of the dark sector fulfill approximately with $T_{_{DE}} \approx T_{_{DM}}$ \cu{at late times}. This assumption, although not general \cite{Pereira:2008at,Poitras:2013deg}, will be used throughout this work. From the total conservation equation (\ref{DE+DM}) \co{and $\dot H$ from (\ref{yy}), the equation (\ref{s3}) is rewritten as} 
\begin{eqnarray}
\dot {S}_{_{DE}}+\dot {S}_{_{DM}}=-\frac{8 \pi L^2}{T}  \left(\dot L-HL \right)\dot H . \label{s4}
\end{eqnarray}
The discussion of negative entropy or negative temperature has been addressed elsewhere \cite{Jamil:2009eb} and has been proposed as a reason to consider the GSL.\\
At this stage, we have to connect the temperature of
the fluids $T$, which is considered moreover equal to that of the horizon $T_h$. This assumption has been made for specific horizon in the literature and we extend this to generalize it to any surface related to the dark energy as it is the case in holographic contexts. Its associated temperature is given as
$T_h \equiv \frac{1}{2 \pi L}$, according to \cite{Jamil:2009eb} and references therein. The entropy associated with the horizon is 
$S_h \equiv 8 \pi ^2 L^2 $.  

In order to obtain variation in the total entropy we add the entropies change rate (\ref{s4}) with the derivative with respect to cosmic time of the entropy $S_h$
\begin{eqnarray}
\dot S_{tot} &=&\dot {S}_{_{DE}}+\dot {S}_{_{DM}}+\dot S_h %\nonumber \\
%&=&
= 16 \pi ^2 L \dot L -16 \pi^2  L^3 \left( \dot L-HL\right)\dot H, \label{st}
\end{eqnarray}
We consider $T_h$ as the temperature of the sources inside the horizon, which is in equilibrium with the temperature associated with the horizon for late times.
Given that the length scale $L$ is positive, we can simplify  (\ref{st}) an thus obtain the total entropy change of the interacting system as
\begin{eqnarray}
\frac{\dot S_{tot}}{16 \pi ^2 L}= \left(1-\dot H L^2\right) \left( \dot L -HL \right) + H L. \label{St}
\end{eqnarray}
The sign of this expression depends only on $(H,L,\dot H,\dot L)$ and is valid regardless of the horizon chosen for the model of Universe considered here. The interacting term is not present nor has played any role in the calculations preceding this result. Eq. (\ref{St}) includes other results where the horizon (and Ansatz) were selected before testing the GSL: in \cite{Karami:2010zz} for apparent and event horizons; in \cite{Bhattacharya:2010zf} for the event horizon; and then in \cite{Bhattacharya:2011} for a generalized RHDE model, also for both horizons.\\
On the other hand, using  (\ref{yy}) for $\dot H$ into  (\ref{St}), we obtain 
\begin{eqnarray}
\frac{\dot S_{tot}}{8 \pi ^2 L^3 H^2 } 
 =\left( 3+\frac{p}{H^2} \right) \left( \dot L -HL \right) +  \frac{2\dot L}{H^2L^2} , 
\label{Sst}
\end{eqnarray}
which depends on $(H,L,p, \dot L)$, where $\dot L$ and $p$ could determine the sign of the entropy change. If we consider that the horizon is not a decreasing function, as expected for an expanding Universe, its derivative is positive and the range where GSL is respected will depend on the total pressure and how negative it can be. We will consider some specific known scenarios and their generalization in the following section.
 
\section{Horizons and Dark Energy} 
In flat FRW, if the horizon is related to the length scale $L$, the dark energy proposed by \cite{Li:2010cj,Li:2004rb} is written as
\begin{eqnarray}
\rho_{_{DE}}\equiv \frac{3c^2}{L^2}, \label{L}
\end{eqnarray}
where $c$ is a positive constant that fulfill ${c}^{2}<1$ \cite{Zimdahl:2002zb,Radicella:2010vf,Zhang:2013mca}. In this context, we can express the coincidence parameter (\ref{r}) as a function of the length scale and the Hubble parameter
\begin{eqnarray}
r=\frac{H^2 L^2}{c^2}-1. \label{rHLc}
\end{eqnarray}
In order to test the validity of GSL we focus on known functions, thus we obtain the length scale from (\ref{rHLc}) in terms of $r$ and its derivative. Then we focus on one expression from (\ref{St})
\begin{eqnarray}
\left(\dot L-HL\right)=\frac{c^2}{L H}\left[\frac{r'}{2}+q(1+r)\right], \label{HrHpuntorprima}
\end{eqnarray}
where we substitute $\dot H$ in terms of the deceleration parameter $q\equiv-\left(1+\frac{\dot H}{H^2}\right),$ whose sign is negative at late times.
If $q<0$ and, for a decreasing coincidence parameter, $r'<0$, we conclude that the parenthesis $(\dot L-HL)$ in (\ref{HrHpuntorprima}) is negative at late times. This will describe a universe with late acceleration and a coincidence parameter that decreases during its evolution. Comparing this with  (\ref{s4}) we notice that the sign of the change of entropy of the matter content will depend only on $\dot H$. Substituting $\dot H$ from the definition of the deceleration parameter and (\ref{HrHpuntorprima}) into (\ref{St}) we obtain
\begin{eqnarray}
\frac{H\dot S_{tot}}{16 \pi ^2 c^2} &=&  \left[1+c^2(1+r)(q+1)\right] \left[ \frac{r'}{2}+(1+r)q\right]  %\nonumber \\
%&& 
+(1+r). \label{S21}
\end{eqnarray}
If we require GSL \co{to be fulfilled on the right side of} (\ref{S21}), the range for the variables $(r,q,r')$  is
\begin{eqnarray}
 -c^2(q+1)>\frac{1}{1+r}+\frac{1}{r'/2+(1+r)q}, \label{qc}
\end{eqnarray}
\co{for a negative $q$ at late times}. The upper or lower limit of the holographic parameter $c^2$ will depend on the deceleration parameter and whether $q>-1$ or $q<-1$. The fulfillment of the GSL and the cosmic acceleration are thus connected in this holographic context, as shown by (\ref{qc}). This inequality has been used to evaluate GSL to obtain an upper limit on $q$ using a specific horizon and an Ansatz for the interaction \cite{Setare:2007at}. \\
Focusing on the entropy change and the interaction, we could use another approach. \co{Using (\ref{L}) and its derivative in (\ref{DE}),} we obtain $\dot L$ as a function of the interaction 
\begin{eqnarray}
\dot L= \frac{HL}{2} \left[3+\frac{L^2}{c^2}\left(p_{_{DE}}+\frac{Q}{3 H} \right) \right], \label{HH}
\end{eqnarray}
Replacing this in  (\ref{Sst}) for a pressureless dark matter component \co{($p_{_{DM}}=0$)} we obtain
\begin{eqnarray}
\frac{S'_{tot}}{4 \pi ^2 L^2} =& 4 + \left(2+3H^2L^2+p_{_{DE}} L^2\right)  %\nonumber \\
% &
\times \left[1+\frac{L^2 }{3c^2}\left(3 p_{_{DE}}+\frac{Q}{ H} \right) \right]. \label{StHLQP}
\end{eqnarray} 
In addition, we consider the coincidence parameter (\ref{rHLc}) into  (\ref{StHLQP}) and we rewrite the total entropy change in terms of $r$ as 
\begin{eqnarray}
\frac{ S'_{tot}}{4 \pi ^2  L^2} 
=& 4+ \left[\left( 3+P_{_{DE}} \right) c^2 (1+r)+2\right] %\nonumber \\
 %&
 \times  \left[  1+(1+r) \left( P_{_{DE}}+\Gamma \right)  \right]  , \label{Scr}
\end{eqnarray}
a function of $(r,P_{_{DE}},\Gamma)$, using $P_{_{DE}}=\frac{p_{_{DE}}}{H^2}$ and $\Gamma=\frac{Q}{3H^3}$. We shall refer to both variables as the interaction.If GSL is required, the right side of equation (\ref{Scr}) must be positive, which leads to inequalities in terms of the interaction according to the sign of each individual term.\\
{We introduce} two auxiliary \co{functions} 
\begin{eqnarray}
&& f(a) \equiv -P_{_{DE}}-\frac{1}{(1+r)}, \nonumber \\
 &&
%\qquad
 g(a)\equiv \frac{-4}{(1+r)\left(\left( 3+P_{_{DE}}\right) c^2 (1+r)+2\right)}, \label{estrella}
\end{eqnarray}
whose sign can vary with time. For instance, for barotropic pressures, $p_{_{DE}}=\omega_{_{DE}} \rho _{_{DE}}$, the sign of $f(a)$ will depend on whether $\omega_{_{DE}}(t) \lessgtr -1/3$. \\ 
 Then, using the functions from (\ref{estrella}), we can rewrite (\ref{Scr}) as
\begin{eqnarray}
\frac{ S'_{tot}}{4 \pi ^2  L^2}
=  -\frac{4}{g(a)}\left( \Gamma-f(a) -g(a) \right), \label{Qstar}
\end{eqnarray}
If GSL is required, the right side of equation (\ref{Qstar}) must be positive, which leads to inequalities in terms of the interaction where the horizon does not play any role. 
 Considering the sign of expression (\ref{HrHpuntorprima}) into this equation we already know that $(\Gamma-f(a))$ is negative. Thus, all possible remaining scenarios are: \\
If $g(a) >0$, then we obtain
\begin{eqnarray}
	 \Gamma< f(a)<f(a)+g(a), \label{28}
\end{eqnarray}
a \co{dynamical} upper limit on the interaction. 
On the other hand if $g(a) <0$, then we obtain
\begin{eqnarray}
	 f(a)> \Gamma>f(a)+ g(a), \label{29}
\end{eqnarray}
an upper and lower dynamical limit to the interacting function. It should be noted that if $f(a)<0$, then the interaction between the fluids in the dark sector has to be negative also in order for the model to respect GSL and alleviate cosmic coincidence at late times. These inequalities do not depend explicitly on the length scale, given that $f(a)$ and $g(a)$ defined in (\ref{estrella}) do not.\\
As an application, we consider a specific horizon related to holographic dark energy (\ref{L}) and study the implications described above for when the GSL is satisfied. Firstly, we shall consider known horizon and compare this approach with results described in the literature and then we shall use the Ricci and Ricci-like proposal.

\subsection{Known horizon and the GSL}
\co{There are known horizon in the literature for holographic dark energy scenarios, we shall analyze briefly the GSL on three of them: the Hubble horizon $L_H$, the event horizon $L_E$ and a linear combination of both $L_{HE}$, where the sub-index will denote the horizon used for the three cases.}\\
The Hubble horizon $L_H$ in the FRW flat universe is given by
\begin{eqnarray}
L_H\equiv H^{-1}, \label{LsubH}
\end{eqnarray}
which is the same as the apparent horizon for a flat configuration. Considering $L_H$ as the length scale, the dark energy for this case is $\rho_{_{DE}}=3c^2 H^2$. Then Eq. (\ref{St}) is given as  
\begin{eqnarray}
 H\frac{\dot S_{H}}{16 \pi ^2 } = \left(H^{-2} \dot H  \right)^2,
\end{eqnarray}
where the interaction plays no role in the sign of $\dot S_{H}$, which is always defined positive regardless of the number of components. Thus the GSL of thermodynamics is fulfilled in a region enclosed by the Hubble horizon throughout the evolution of the Universe.  
This result is a particular case of the work of \cite{Setare:2008bb,Setare:2009ti} for a flat model. \\
For the horizon (\ref{LsubH}) the cosmic coincidence parameter (\ref{r}) is given as
$r_H=
\frac{1}{c^2}-1,$
a positive constant and a scaling regime, considering that $c^2<1$ \cite{Zhang:2013mca}. This HDE therefore does not alleviate the cosmic coincidence problem. \\
The cosmological event horizon $L_E$ is defined as
\begin{eqnarray}
L_E\equiv a \int _t^{\infty}\frac{dt}{a}=a \int _t^{\infty}\frac{da}{H a^2}, \label{LE}
\end{eqnarray}
therefore the derivative of the event horizon with respect to time simplifies as 
$\dot L_E=HL_E-1.$
 Considering the event horizon (\ref{LE}) and its derivative in  (\ref{St}), we obtain an expression that depends on $(H,L_E,\dot H)$. Using the coincidence parameter defined in (\ref{r}) we obtain
\begin{eqnarray}
\frac{\dot S_{E}}{16 \pi ^2 L_E}= 2c^2(1+r)^2 r'-c^2(1+r)-1.
\end{eqnarray}
This expression depends on $(r,r')$; if $r$ is a decreasing function over late times, then $r'$ is a negative function and the GSL is not respected. No interaction or variable EoS could have change that result. This result coincides with the work of authors in \cite{Debnath:2010za} for their chosen Ansatz and it was briefly mentioned at the end of \cite{Jamil:2009eb}. \\
Now, we consider a linear combination of the Hubble horizon $L_H$ and the Event Horizon $L_E$ assumed by \cite{Sadjadi:2006qb}, while investigating thermodynamics and the phantom barrier. This is given as
\begin{eqnarray}
L_{HE}\equiv \lambda_H L_H+ \lambda_E L_E, \label{LsubHE}
\end{eqnarray}
where $L_H$ and $L_E$ are defined in (\ref{LsubH}) and (\ref{LE}) respectively, and $\lambda_H$ and $\lambda_E$ are both positive constants. 
The derivative of (\ref{LsubHE}) with respect to time simplifies as 
$\dot L_{HE}=HL_{HE}+\lambda_H-\lambda_E.$ 
This expression \co{is not always positive, as can be inferred comparing it with (\ref{HrHpuntorprima}), where we conclude that $\lambda_H<\lambda_E$, thus $\dot L_{HE}$ could be negative at some point}. \\
Considering this horizon as the length scale and expressing the total entropy change in terms of the coincidence parameter defined in (\ref{r}) and its derivative, 
we can obtain $\dot H$ in terms of $(r,r')$ and replacing this term into (\ref{St}), we get the entropy change
\begin{eqnarray}
\frac{\dot S_{HE}}{16 \pi ^2 L_{HE}}=& \left[1- c^2\frac{ r'}{2}+c^2(r+1)\right] \left(\lambda_H-\lambda_E\right) %\nonumber \\
 %&
+ \left[(\lambda_H-\lambda_E)^2 + 1\right] c\sqrt{r+1} ,
\end{eqnarray}
the GSL can be respected conditioned by the evolution of the parameters.  \\
Of the three horizon analyzed, one of them does not alleviates the cosmic coincidence problem, the second does not respect the GSL and the third could, but only for a certain range. We shall move on to more general scenarios to describe Ricci and Ricci-like holographic dark energy density.

\subsection{Ricci and Ricci-like scenarios}
We consider that since the dark energy is related to the horizon in an holographic context, and Ricci and Ricci-like scenarios consider the dark energy as an specific Ansatz, we shall study the possibility of using such Ansatz as an expression for a surface representing the horizon in a thermodynamical context. 
The Ricci dark energy density is
\begin{eqnarray}
\rho_{_{DE}}\equiv 3 c^2 (2 H^2+\dot{H}) , \label{HL}
\end{eqnarray}
where $c^2$ is a positive constant as proposed by \cite{Gao:2007ep}. This formulation considering an Ansatz on the dark energy density \co{for $L^{-2}=2H^2+\dot H $} is equivalent, to an Ansatz for the dark energy pressure given as
\begin{eqnarray}
p_{_{DE}}=H^2-\frac{2}{3 c^2}\rho_{_{DE}}, \label{palpha}
\end{eqnarray}
\co{obtaining $\dot H$ from (\ref{HL}) and (\ref{yy}) with} pressureless dark matter component $p_{_{DM}}=0$. In \cite{Nojiri:2005pu} there is an expression for the pressure as a function of $p=p \left(H^2,\dot H \right)$. This association differs from the results from the previous section and is a consequence of the $\dot H$ term within the Ansatz and therefore present in the length scale. 
 Using $\rho_{_{DE}}$ from (\ref{HL}) into (\ref{L}), we can obtain $\dot H$,
 therefore, \co{using (\ref{rHLc})} in this model, the deceleration parameter is
$q= 1-1/(  c^2 (1+r)) $. Furthermore, if we replace the dark energy pressure (\ref{palpha}) and (\ref{L}) in (\ref{HH}) we obtain 
$\dot L = \frac{HL}{2 c^2} \left(3c^2-2+{H^2L^2}\left(1+\Gamma \right) \right)$. This differs from the previous section because $\dot L$ has a direct dependence on the interaction, which was not the case in \co{the known horizons}.
Replacing $\rho_{_{DE}},\dot L$ \co{and $p_{_{DE}}$ from (\ref{L}) and (\ref{palpha})} into  (\ref{St}) results in $\dot S_{R}$ depending on $(H,L,\Gamma )$, then \co{using} (\ref{rHLc}) we obtain
\begin{eqnarray}
 \frac{H\dot S_{R}}{8 \pi ^2 c^2 (1+r) } &=&2c^2(1+r)\left[1+(r+1)\left(1+\Gamma\right) \right] %\nonumber \\
%&&
- 2(1+2r) \label{48}
\end{eqnarray}
equation \co{that} depends on $(r,\Gamma)$. This result resembles \cite{Bhattacharya:2011} for its non interacting model explicit in the \co{redshift}, $H$ and $L$. 

\co{Using $\rho_{_{DE}}$ in terms of $r$ and }the pressure (\ref{palpha}), \co{we obtain} the \co{functions} (\ref{estrella}) given as $f(a)=-1+\frac{2-c^2}{c^2(1+r)}$, $g(a)=-\frac{1}{c^2(1+r)^2}$, where we see that the auxiliary \co{function} $g(a)$ is negative. The latter implies that the inequality (\ref{28}) is discarded and we are left with only (\ref{29}), an upper and lower limit, therefore if $c^2>\frac{2}{2+r}$, then interaction could be negative, on the other hand if $c^2<\frac{(1+2r)}{(2+r)(1+r)}$, then the right side of inequality (\ref{29}) is always positive and therefore, the interaction $\Gamma$ should { always be positive} at late times. 

Now, we consider the \co{entropy change} without the interaction \co{explicitly, but {rather}} in terms of $r$ and $r'$. For instance $(\dot H,\rho_{_{DE}})$ can always be written as a function of \co{$(L,H)$ using (\ref{r}), (\ref{L}) and (\ref{palpha})}. With this we can use the derivation \co{of} $L^2=c^2\frac{1+r}{H^2}$ and obtain $\dot L$ \co{and $\dot H$} written as a function of $(r',r)$.
Therefore,  (\ref{St}) is given as
\begin{eqnarray}
\frac{H \dot S_{R}}{16 \pi ^2 } &=& \left[1-\frac{1}{c^2} + 2(1+r)\right]  \left[ \frac{r'}{2}-\frac{1}{ c^2 } + (1+r)\right] %\nonumber \\
%&&
 +(1+r),\label{Strp}
\end{eqnarray}
an expression depending only on $(r,  r')$ and equivalent to (\ref{48}). Studying an Ansatz on $r$ would allow to know if the system respects GSL through this equation.

We recall that the interaction can be written in terms of $p_{_{DE}}$ and  $r$ according to (\ref{Qpr}); however in this context by choosing a horizon we are inherently choosing a pressure. 
Using the pressure (\ref{palpha}) \co{we obtain} $P_{_{DE}}$ as a function of $r$ and $c^2$
\begin{eqnarray}
P_{_{DE}}=-\left[\frac{2}{ c^2 (1+r)} -1  \right], \label{prc}
\end{eqnarray}
The result obtained for the pressure (\ref{prc}) indicates that the evolution of $r$ could be determined by the evolution of $P_{_{DE}}$, relation obtained from the holographic context in (\ref{HL}). 
Then the interaction (\ref{Qpr}) can be written as
\begin{eqnarray}
 \Gamma =\frac{  r'+r \left[\frac{2}{c^2} -  (1+r) \right]}{ (1+r)^2} ,  \label{Qrc}
\end{eqnarray}
a function\footnote{Let us note that the interaction is null for a particular Ansatz of the coincidence parameter 
$r=\frac{ (c^2-2) }{ a^{(2/c^2-1) } (c^2(1+r_0)-2)/r_0- c^2}$.} of $(r,r')$. The interaction $\Gamma$ has been formulated as a function of derivatives of $r$ before in 
\cite{Arevalo:2011hh} and references therein, but that result differ from ours because the EoS is given as a constant bound to cross the phantom barrier that determines the sign of $\Gamma$. In our case the sign of  $\Gamma$ remains undetermined so far.\\
Using $r$ as a function of $P_{_{DE}}$ from (\ref{prc}) and its derivative in the interaction (\ref{Qpr}), we obtain
\begin{eqnarray}
\Gamma=\frac{c^2}{2 } \left[ P'_{_{DE}}-P_{_{DE}}\left(\frac{2}{c^2}-1+ P_{_{DE}} \right)\right],  \label{Qpc}
\end{eqnarray}
which depends on $\left(P_{_{DE}},   P'_{_{DE}}\right)$.

\co{Considering $0<r<1$, then we can plot the phase-space of $(r,c^2)$ for Ricci Dark Energy in Figure 1, }
where we see that the shaded area for a negative pressure almost excludes negative only interactions. The possibility of sign change in \co{$\Gamma$} for different stages of evolution is allowed for several ranges and \co{that} $c^2$ and $r$ are bounded {in order to do so}. Moreover, this has been studied for a given interaction, thus we can compare those Ansatz with the GSL in the phase-space. The pair of dots $(r_0,c^2)$ respect the GSL and are inside the $\Gamma>0$ region, as suggested by the observational data of interaction between two fluids, estimated to be a positive quantity today \cite{Salvatelli:2014zta,Cai:2015zoa}.

\begin{figure}[htbp]
	\centering
		\includegraphics[width=0.5\textwidth]{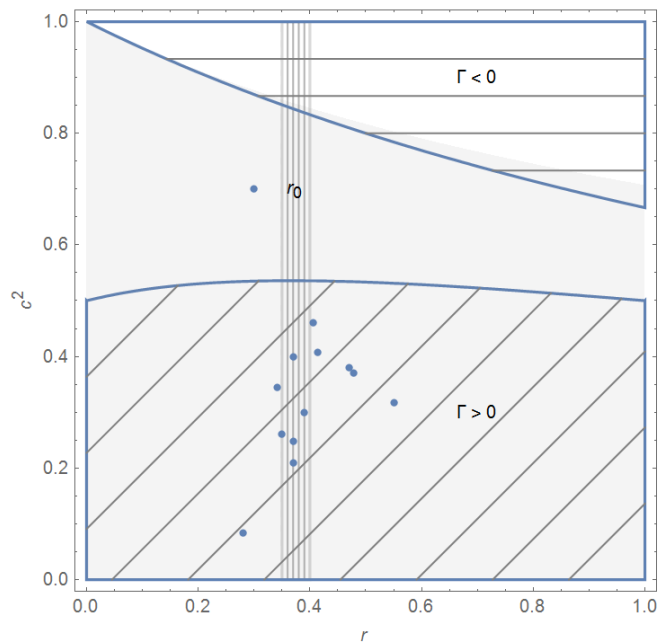}
		\caption{Phase-space of $(r,c^2)$ considering zones where the interaction $Q$ respects the GSL for Ricci Dark Energy. The shaded area is where the pressure is negative at late times. The dots are pairs of $(r_0,c^2)$ obtained with observational data from \cite{Gao:2007ep} and references therein, and the vertical lines for $r_0$ are indicated for comparison.}
\end{figure}

We investigate the case of the generalization of Ricci dark energy (\ref{HL}), the Ricci-like scenario. The Ricci-like dark energy density is given by
\begin{eqnarray}
\rho_{_{DE}}\equiv 3(\alpha H^2+\beta\dot{H}) , \label{HRL}
\end{eqnarray}
where $\alpha$ and $\beta$ are both constant as proposed by \cite{Granda:2008dk} and for non-interacting cosmic fluids both constants are positive and less than one according to the comparison performed by \cite{Lepe:2010vh,Granda:2011zm,Fei:2013oea}. The main difference is the number of parameters, nevertheless, Ricci-like contains the previous case for the equivalence $(\alpha \rightarrow 2c^2,\beta \rightarrow c^2)$. Henceforth, we consider in  (\ref{L}) the constant $c$ is included in the respective constants for each horizon.  Additionally, interacting Ricci-like scenarios present a change of sign in the interaction, not described in Ricci interaction \cite{Arevalo:2013tta}. Comparing with (\ref{L}) we obtain a relation between the horizon and the Hubble parameter and its derivative {given as $L^{-2}=\alpha H^2+\beta\dot{H}$.} This Ansatz (\ref{HRL}) has an implicit equivalence with other models for the dark energy pressure described in \cite{Capozziello:2005pa}, which can be clearly seen obtaining $\dot{H}$ from the Ansatz (\ref{HRL}) and replacing it into  (\ref{yy}) for a pressureless dark matter configuration, as
\begin{eqnarray}
p_{_{DE}}=-\frac{2}{3 \beta} {\rho}_{_{DE}}+(2 \alpha-3\beta) \frac{H^2}{\beta}. \label{pDE}
\end{eqnarray}
This is the generalization of the pressure for RDE (\ref{palpha}).
The expression for (\ref{pDE}) is not polytropic but resembles inhomogeneous fluid cosmology, where the pressure is dependent not only on the dark energy density but also on the Hubble parameter, \co{as noted by \cite{Arevalo:2013tta}}. This is caused by the $\dot H$ term in (\ref{HRL}) as was the case in the Ricci Ansatz.\\
By considering the definition of the deceleration parameter and $\dot H$ from (\ref{HRL}) as a function of $(H,L)$, we obtain 
$q =\frac{1}{\beta }\left( {\alpha  - \beta  - \frac{1}{1 + r }} \right)$, whereas if we have an Ansatz for the coincidence parameter $r(a)$, the deceleration parameter $q(a)$ can be obtained, and thus we can study whether the model is accelerated in late times. 
It should be noted that for a constant $r$, the deceleration parameter $q$ would also be constant, thus there would be no evolution from a non-accelerated Universe to an accelerated one. \\
With the pressure (\ref{pDE}) and (\ref{L}) we can rewrite equation (\ref{HH}) as 
$%\begin{eqnarray}
\dot L = \frac{HL}{2} \left[3-\frac{2}{ \beta } + H^2L^2\left(\frac{2 \alpha-3\beta}{\beta}  +\Gamma \right)\right]
$. %\end{eqnarray} 
 \cu{Substituting} this into (\ref{Sst}) along with the pressure and energy density from (\ref{pDE}) and (\ref{L}) respectively, we obtain 
an equation \co{for the change of entropy} that depends on $(H,L,Q)$.
Considering this equation in terms of the coincidence parameter\co{, using (\ref{rHLc}),} we obtain
\begin{eqnarray}
H\frac{\dot S_{Rl}}{8 \pi ^2 (1+r)}= 2+\left[1-\frac{1}{\beta }\left(1-\alpha (1+r) \right) \right] %&& \nonumber \\
 \times\left[1-\frac{2}{ \beta } + (1+r)\left(\frac{2 \alpha-3\beta}{\beta}  +\Gamma \right)\right], && \label{rQ}
\end{eqnarray}
an expression that depends on  $(r,\Gamma)$. The sign on the first parenthesis will depend on the relation between $\alpha$, $\beta$ and $r$; moreover the second parenthesis 
is negative at late times, independent of the horizon.

For the pressure (\ref{pDE}) the \co{functions} (\ref{estrella}) are
$f(a) =\frac{2-\beta}{\beta(1+r)}-\frac{2\alpha-3\beta}{\beta}, \ g(a)=\frac{-2 \beta /(1+r)}{\beta-1+\alpha(1+r)}
$. The sign of interaction in Ricci-like HDE can be negative or positive while respecting GSL, according to the ranges obtained from replacing $(f,g)$ in (\ref{28}) and (\ref{29}), that are inequalities in terms of $r$, $\alpha$ and $\beta$.

Given that we do not yet know the sign of the interaction for the Ricci-like scenario we attempt a different approach in order to obtain an expression independent of $\Gamma$. 
Rewriting $\dot H$ in terms of $r$ and $r'$ instead of $\Gamma$ or $p_{_{DE}}$, 
the dark energy density (and thus $L$) can always be written as a function of $(r,H)$ as $L^2 =\frac{1+r}{H^2}$. 
Therefore, we obtain
\begin{eqnarray}
\beta ^2 H  \frac{\dot S_{Rl}}{16 \pi ^2 } = \beta^2  (r+1)+ \left[\beta-\left(1 - \alpha ( r+1)\right)  \right] 
 %\nonumber \\
 \times \left[\frac{\beta  }{ 2  }r' - 1+(\alpha -\beta ) (1+r) \right], \label{Qrrprime}
\end{eqnarray}
an expression depending only on $(r,  r')$. We can observe that studying an Ansatz on $r$ would  immediately allow us to analyze whether this Ansatz respects the GSL, as in the previous case of (\ref{Strp}).  

We recall that the interaction can be written in terms of $p_{_{DE}}$ and  $r$ according to (\ref{Qpr}); however in this context by choosing a horizon we are inherently choosing a pressure. 
Using the pressure (\ref{pDE}) \co{we obtain} $P_{_{DE}}$ as a function of $r$ and the model parameters
\begin{eqnarray}
P_{_{DE}}=-\frac{1}{\beta}\left[\frac{2}{  (1+r)} -(2 \alpha-3\beta)  \right], \label{pr}
\end{eqnarray}
The result obtained for the pressure (\ref{pr}) indicates that the evolution of $r$ could be determined by the evolution of $P_{_{DE}}$, relation obtained from the holographic context in (\ref{HRL}). Then the interaction (\ref{Qpr}) can be written as
\begin{eqnarray}
 \Gamma =\frac{ \beta \ r'+r \left(2 -(2 \alpha-3\beta) (1+r) \right)}{\beta (1+r)^2} 
,  \label{Qr}
\end{eqnarray}
a function of $(r,r')$. \footnote{Let us note that the interaction is null for a particular Ansatz of $r=\frac{ r_0(2 \alpha-3\beta-2) }{ a^{(2-2 \alpha+3\beta)/\beta }( (2 \alpha-3\beta)(1+r_0)-2)-r_0(2 \alpha-3\beta)}$.}
\\
Using $r$ as a function of $P_{_{DE}}$ from (\ref{pr}) and its derivative in the interaction (\ref{Qpr}), we obtain
\begin{eqnarray}
\Gamma=\frac{1}{2} \left( \beta P'_{_{DE}}-P_{_{DE}}\left(2(1-\alpha)+\beta(3+ P_{_{DE}}) \right)\right),  \label{Qp}
\end{eqnarray}
which depends on $\left(P_{_{DE}},   P'_{_{DE}}\right)$. In the expressions (\ref{Qr}) and (\ref{Qp}), it can be observed that the sign of the interaction can change according to the choice of Ansatz ($P_{_{DE}}$ or $r$) and the sign of its respective derivatives. It is equivalent to study the expression for interaction described by (\ref{Qr}) as it is the interaction as a function of the pressure $P_{_{DE}}$. 

The cosmological scenario, for an Ansatz in the coincidence parameter given as $r=r_0 a^{-3 \epsilon}$, was analyzed with observational data from Supernovae Ia and the results are presented in Figure 2 (left), obtaining  $\epsilon_{min}=0.169$. The interaction presents a change of sign before the Universe goes from a decelerated to an accelerated phase, however the total entropy change remains positive until the universe reaches 1.5 times its current size, thus respecting the GSL. This can also be seen in the parameter space of ($\alpha,\beta$) for $r_0=0.318$ in Figure 2 (right), where it can be negative in a certain region of this space. This negative interaction region that can respect GSL is introduced due to considering the HDE as the radius from the surface representing the horizon. 

\begin{figure}[htbp]
	\centering
		\includegraphics[width=0.52\textwidth,keepaspectratio]{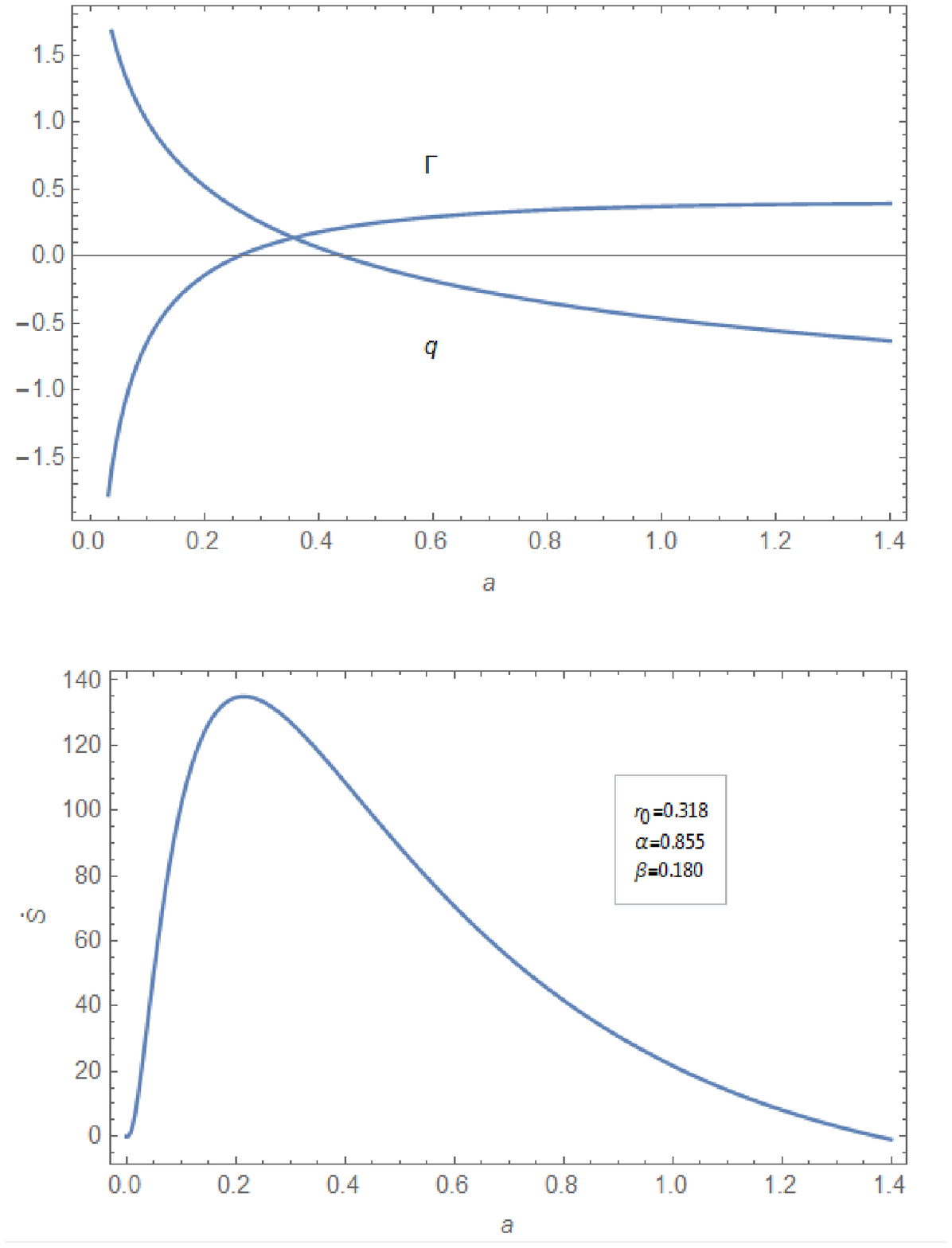}
		\includegraphics[width=0.52\textwidth]{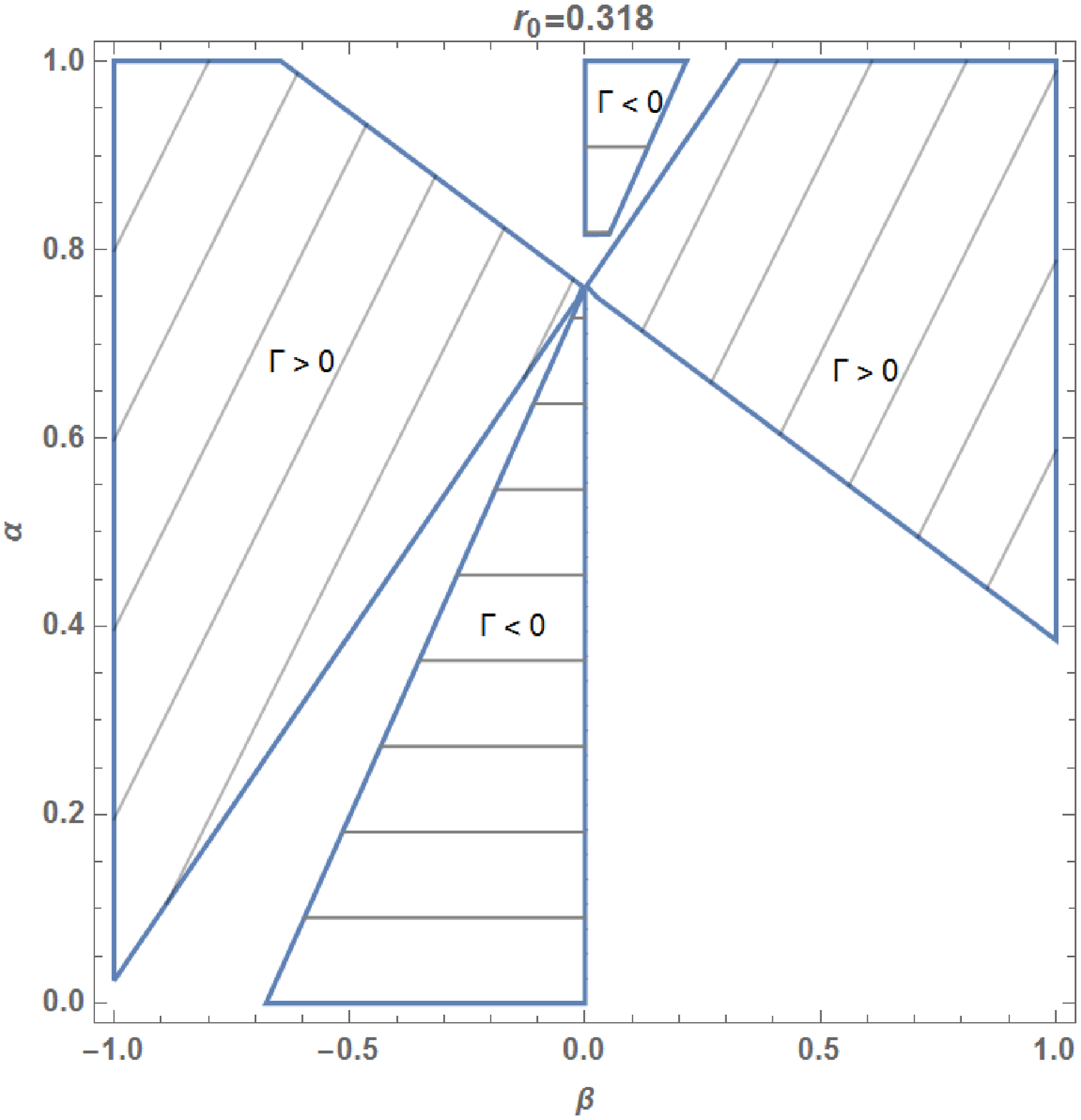}
		\caption{The graph on the bottom represents {the relation between the GSL and} the interaction for a specific $r_0$ in the phase space of ($\alpha,\beta$) and the graph on the center is the entropy change and the top is the deceleration parameter and interaction for a specific holographic model, case III of \cite{Arevalo:2013tta} .}
	\label{Figura1}
\end{figure}

\section{Discussion}
The novelty of our work resides in the fact that we propose the HDE Ansatz as an horizon\co{-like} radius to study features of the cosmological interaction using thermodynamics. Furthermore, we consider the interaction with an unknown sign, generalizing previous studies where the interaction is generally given and positive.

In this paper, we studied the effect of GSL in a flat FRW universe at late times with holographic dark energy interacting with non-relativistic dark matter. The interaction $Q$ is not a phenomenological
function given \textit{a priori}, but an unknown variable to be studied according to the thermodynamics of the system in equilibrium conditions at late times. In this context, we examined the validity of the GSL of thermodynamics considering the universe as a thermodynamical system bounded by the length scale $L$.
We see from (\ref{Qpr}) that besides the possibility of a change in the \cu{sign} of $Q$ interaction; this equation is independent of the choice of horizon but depends on the coincidence parameter, its derivative and the EoS. Considering an unknown $Q$, we analyze whether this interacting model respects the GSL of thermodynamics. 

When choosing the holographic dark energy density (\ref{L}), one can obtain expressions for the change in total entropy in terms of the deceleration parameter (\ref{S21}). The GSL has implications for the cosmic acceleration of the universe; from (\ref{qc}) it can be seen that the holographic parameter $c^2$ has an upper or lower limit depending on whether the deceleration parameter is greater or less than $-1$, respectively. 
When we applied the GSL of thermodynamics of a surface of radius $L$ in HDE (\ref{Qstar}), we obtain two ranges determined by the dark energy pressure, (\ref{28}) and (\ref{29}), where  the unknown interacting function can respect this law. Regarding these ranges, we can conclude that for a negative $f(a)$, the interaction can take positive or negative values, or a transit between the two options and within a certain range respect the GSL of thermodynamics. The range obtained is dynamic as a consequence of the HDE and a negative interaction in this context cannot be ruled out.

 To examine further the validity of the GSL, we considered different radius, three known horizon and two new proposals. Unlike other works, we used the horizon for the holographic dark energy density as surface representing a cosmological horizon. The variation with respect to the cosmic time of the total entropy can be positive or negative depending on the horizon chosen when this horizon includes a term of $\dot H$ that inherently determines the pressure. These results are verified in Ricci and Ricci-like horizons, where there is an equivalence between the coincidence parameter $r$ and the dark energy pressure $p_{_{DE}}$.
For these cases the cosmic coincidence problem is alleviated, given that when the pressure is a variable function, $r$ is also a variable function. For the RDE model, the analysis indicates that the interaction must be mainly positive today in order to respect the GSL, as can be seen from Fig 1. On the other hand, the Ricci-like HDE is allowed to be negative and respect GSL but only for a certain range, noticeable in Fig 2 (bottom). The viability of this model can be seen in the observational contrast presented in Fig 2 (top and center).

Overall, we integrate and generalize previous works, obtain new results and connections among model, thus expanding the studies on holographic dark energy and sign-changeable interactions.
The aspects introduced by each model at different phases in the evolution of the Universe will be studied in future research.

\section*{Acknowledgments}
This work is dedicated to the memory of our colleague Sergio del Campo, for his contributions to Cosmology in Chile. 
This work has been supported by Comisi\'on Nacional de Ciencias y Tecnolog\'ia
through Fondecyt Grants 3130736 (FA).
 (FP) acknowledges DI14-0007 of Direcci\'on de Investigaci\'on y Desarrollo for basal institutional support and  Universidad de La Frontera, institution that also contributes reviewing the English in our work.  (PC) acknowledges support from the Department of Physics of Universidad de La Frontera. (FA) would like to thank A. Cid and P. Mella for helpful references.

%%%%%%%%%%%%%%%%%%%%%%%%%%%%%

\end{document}